%% file: wts11.tex
\documentclass[conference]{IEEEtran}
\ifCLASSINFOpdf
  \usepackage[pdftex]{graphicx}
\else
\fi
\usepackage{amsmath}
\usepackage{amssymb}
\usepackage{amsfonts}
\usepackage{subfig}
\begin{document}
%
\title{Generic Approach for Hierarchical Modulation Performance Analysis: Application to DVB-SH}

\author{\IEEEauthorblockN{Hugo M{\'e}ric\IEEEauthorrefmark{1}\IEEEauthorrefmark{2},
J{\'e}r{\^o}me Lacan\IEEEauthorrefmark{2}\IEEEauthorrefmark{1},
Caroline Amiot-Bazile\IEEEauthorrefmark{3}, 
Fabrice Arnal\IEEEauthorrefmark{4} and
Marie-Laure Boucheret\IEEEauthorrefmark{2}\IEEEauthorrefmark{1}}
\IEEEauthorblockA{\IEEEauthorrefmark{1}T{\'e}SA, Toulouse, France}
\IEEEauthorblockA{\IEEEauthorrefmark{2}Universit{\'e} de Toulouse, Toulouse, France}
\IEEEauthorblockA{\IEEEauthorrefmark{3}CNES, Toulouse, France}
\IEEEauthorblockA{\IEEEauthorrefmark{4}Thales Alenia Space, Toulouse, France\\
Email: hugo.meric@isae.fr, jerome.lacan@isae.fr, caroline.amiot-bazile@cnes.fr,\\
fabrice.arnal@thalesaleniaspace.com, marie-laure.boucheret@enseeiht.fr}}


\maketitle

\begin{abstract}
Broadcasting systems have to deal with channel diversity in order to offer the best rate to the users. Hierarchical modulation is a practical solution to provide several rates in function of the channel quality. Unfortunately the performance evaluation of such modulations requires time consuming simulations. We propose in this paper a novel approach based on the channel capacity to avoid these simulations. The method allows to study the performance in terms of spectrum efficiency of hierarchical and also classical modulations combined with error correcting codes. Our method will be applied to the DVB-SH standard which considers hierarchical modulation as an optional feature.
\end{abstract}


%
\IEEEpeerreviewmaketitle

\input{introduction}

\input{theory}

\input{result}

\input{conclusion}

\nocite{*}
\bibliographystyle{IEEEtran}
\bibliography{biblio}

\end{document}

%% file: introduction.tex
\section{Introduction}

In most broadcast applications, the receivers do not experience the same signal-to-noise ratio (SNR). For instance, in satellite communications  the channel quality decreases with the presence of clouds in Ku or Ka band or with shadowing effects of the environment in lower bands.


The first solution for broadcasting is to design the system for the worst-case reception. However this solution does not take into account the variability of channel qualities. This holds a loss of spectrum efficiency for users with good reception. Then two other schemes have been proposed in \cite{cover} and \cite{bergmans} to improve the first one: time division multiplexing with variable coding and modulation, and superposition coding. Time division multiplexing consists in using a first couple modulation/coding rate during a fraction of time, and then using an other modulation/coding rate for the remaining time. All the population can receive the first part of the signal called the High Priority (HP) signal and only the receivers in good conditions receive the second one called the Low Priority (LP) signal.

\begin{figure}[!ht]
\centering
\includegraphics[width = 0.95\columnwidth]{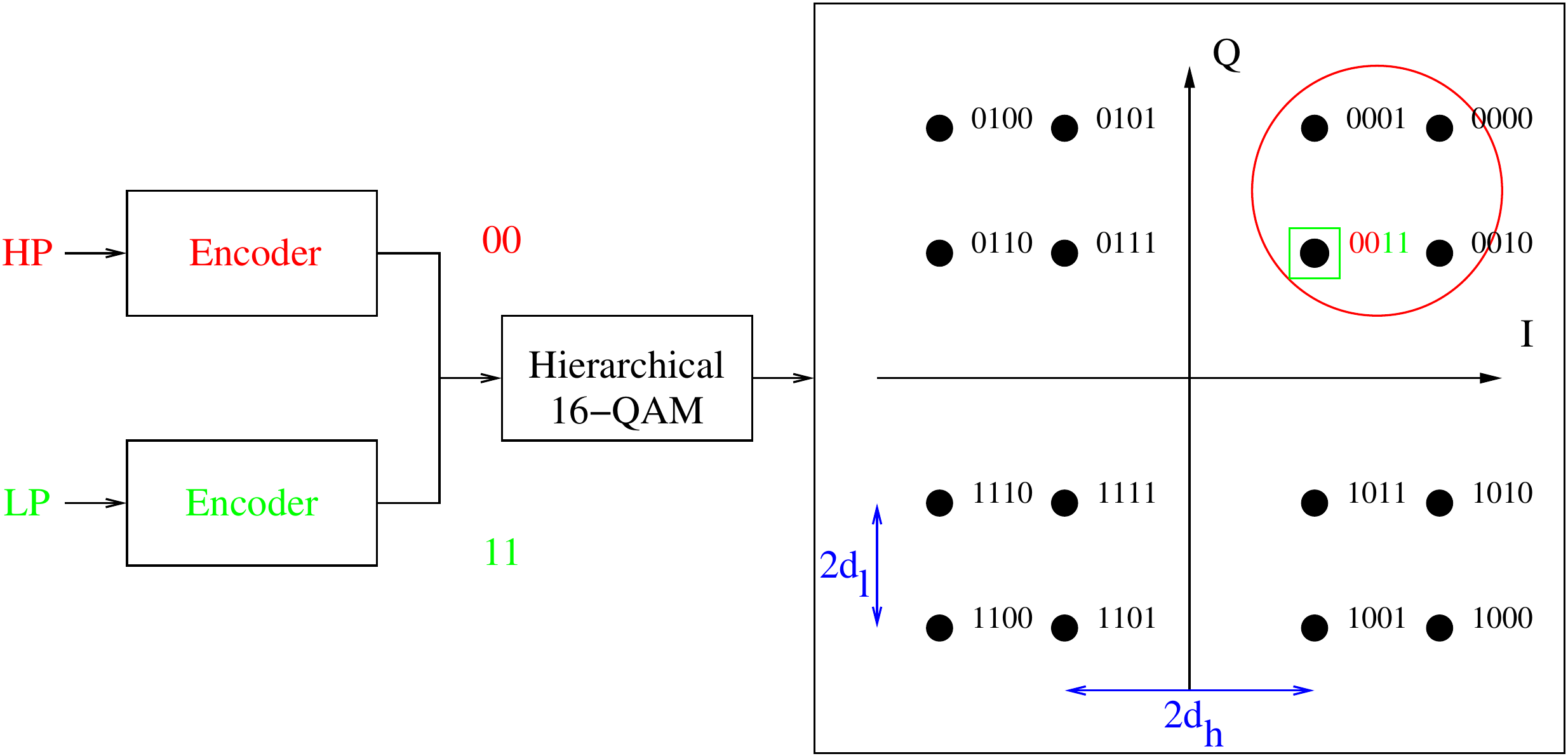}
\caption{Hierarchical Modulation using a non-uniform 16-QAM}
\label{hm_principle}
\end{figure}

Unlike time sharing, superposition coding sends information for all the receivers all the time and it was proved to achieve better rates than time sharing for the continuous Gaussian channel \cite{bergmans}. In superposition coding the available energy is shared to several service flows which are sent simultaneously and in the same band. Hierarchical modulation is a practical implementation of superposition coding. Figure~\ref{hm_principle} presents the principle of the hierarchical modulation using a non-uniform 16-QAM. The idea is to merge two different streams at the modulation step. The HP stream is used to select the quadrant, and the LP stream selects the position inside the quadrant. In good conditions receivers can decode both streams, unlike bad receivers which only locate the quadrant and then decode the HP stream as a QPSK constellation. 

In practical contexts, both schemes can be used combined with applications generating hierarchical flows, like H.264/SVC video \cite{svc_hm,svc_vcm}.

Satellite communications standards like DVB-S2 \cite{s2} and DVB-SH \cite{DVBSH} consider hierarchical modulation as an optional feature. To our knowledge, the only way to estimate the performance of real code using hierarchical modulation is to run simulations. We propose here a new approach to evaluate the performance of such modulation. The method is based on the channel capacity and relies on the fact that the real code at coding rate $R$ works as a theoretical ideal code at coding rate $\tilde R$. Our approach is applied to DVB-SH.

The paper presents our work as follows:
\begin{itemize}
\item In Section~\ref{theory} we compute the capacity for any hierarchical modulation.
\item We propose a method using the capacity to evaluate the performance of hierarchical modulations in terms of spectral efficiency and required $E_s/N_0$ for a targeted Bit Error Rate (BER) or Packet Error Rate (PER) in Section~\ref{result}.
\end{itemize}
Section~\ref{conclusion} concludes the paper by summarizing the results and presents the future work.


%% file: theory.tex
\section{Capacity of the Hierarchical Modulation}\label{theory}

In this section we introduce the channel capacity and compute it for any hierarchical modulation. Some results are given at the end of this section using the hierarchical 16-QAM.

\subsection{Computation of the capacity}
A channel can be considered as a system consisting of an input alphabet, an output alphabet and a transition matrix $p(y|x)$. We define two random variables $X$, $Y$ representing the input and output alphabets respectively. The mutual information between $X$ and $Y$, noted $I(X;Y)$, measures the amount of information conveyed by Y about X. For two discrete random variables $X$ and $Y$, (\ref{defmi}) gives the expression of the mutual information. It can be extended to continuous random variables \cite{cove_thom_91}.
\begin{eqnarray}
I(X;Y) = \sum_{x \in \mathcal{X}} \sum_{y \in \mathcal{Y}} p(x) p(y|x) \log_{2} \left( \frac{p(y|x)}{p(y)} \right).
\label{defmi}
\end{eqnarray}

Using this notion, the channel capacity is then given by
\begin{eqnarray}
C = \max_{p(x)} I(X;Y) ,
\label{mi}
\end{eqnarray}
where the maximum is computed over all possible input distributions.

Here we consider the memoryless discrete input and continuous output Gaussian channel. The discrete inputs $\mathbf{x}_i$ are obtained using a modulation and belong to a set of discrete points $\chi \subset \mathbb{R}^2$ of size $|\chi|=M=2^m$ called the constellation. Thus, each symbol of the constellation carries m bits. An explicit formula (\ref{capacity_berrou}) for the capacity in this particular case is given in \cite[Chapter 3]{berrou}.
\setlength{\arraycolsep}{0.0em}
\begin{eqnarray}
C &{}={}& \log_{2}(M) - \nonumber\\
&&\frac{1}{M} \sum_{i=1}^M \int\limits_{-\infty}^{+\infty} \int\limits_{-\infty}^{+\infty} p(\mathbf{y}|\mathbf{x}_i) \log_{2} \left( \frac{\sum_{j=1}^M p(\mathbf{y}|\mathbf{x}_j)}{p(\mathbf{y}|\mathbf{x}_i)}\right) \, \mathrm d\mathbf{y}
\label{capacity_berrou}
\end{eqnarray}
\setlength{\arraycolsep}{5pt}

We are now interested to evaluate the capacity of a stream using a \emph{subset} of all the bits. Reference \cite{caire} presents the case where each stream uses one bit. The idea is to modify the input random variable $X$ by the input bits used in each stream. We define $b_i$ as the value of the i\textsuperscript{th} bit of the label of any constellation point $\mathbf{x}$. Suppose the stream uses \emph{k bits among m} in the positions $i_1$, ..., $i_k$. For any integer i, let $l_n(i)$ denote the n\textsuperscript{th} bit in the binary representation of i such as $i=\sum_{n=1}^{+\infty} l_n(i)2^{n-1}$. Then using (\ref{defmi}) with the new input variable and continuous output, the capacity of the stream is
\begin{eqnarray}
C = \frac{1}{2^k} \sum_{i=0}^{2^k-1} \int\limits_{-\infty}^{+\infty} \int\limits_{-\infty}^{+\infty} \underbrace{p\left(\mathbf{y}|b_{i_1}=l_1(i),...,b_{i_k}=l_k(i) \right)}_{\text{sum over all possible k-uplets}} \nonumber\\
\log_{2} \left( \frac{ p\left(\mathbf{y}|b_{i_1}=l_1(i),...,b_{i_k}=l_k(i) \right) }{ p(\mathbf{y}) }\right) \, \mathrm d\mathbf{y} ,
\label{capacity_hm}
\end{eqnarray}
where $p(\mathbf{y}) = \frac{1}{2^k} \sum_{i=0}^{2^k-1} p\left(\mathbf{y}|b_{i_1}=l_1(i),...,b_{i_k}=l_k(i) \right)$.

Let $L_n(\mathbf{x})$ denote the n\textsuperscript{th} bit of the label of $\mathbf{x}$. We introduce $\chi_i$ the subset of $\chi$ defined as follows
\begin{eqnarray}
 \chi_i = \{ \mathbf{x} \in \chi | L_{i_1}(\mathbf{x})=l_1(i) , ... , L_{i_k}(\mathbf{x})=l_k(i) \}.
\label{chi}
\end{eqnarray}
 
\begin{figure}[!ht]
\centering
\includegraphics[width = 0.65\columnwidth]{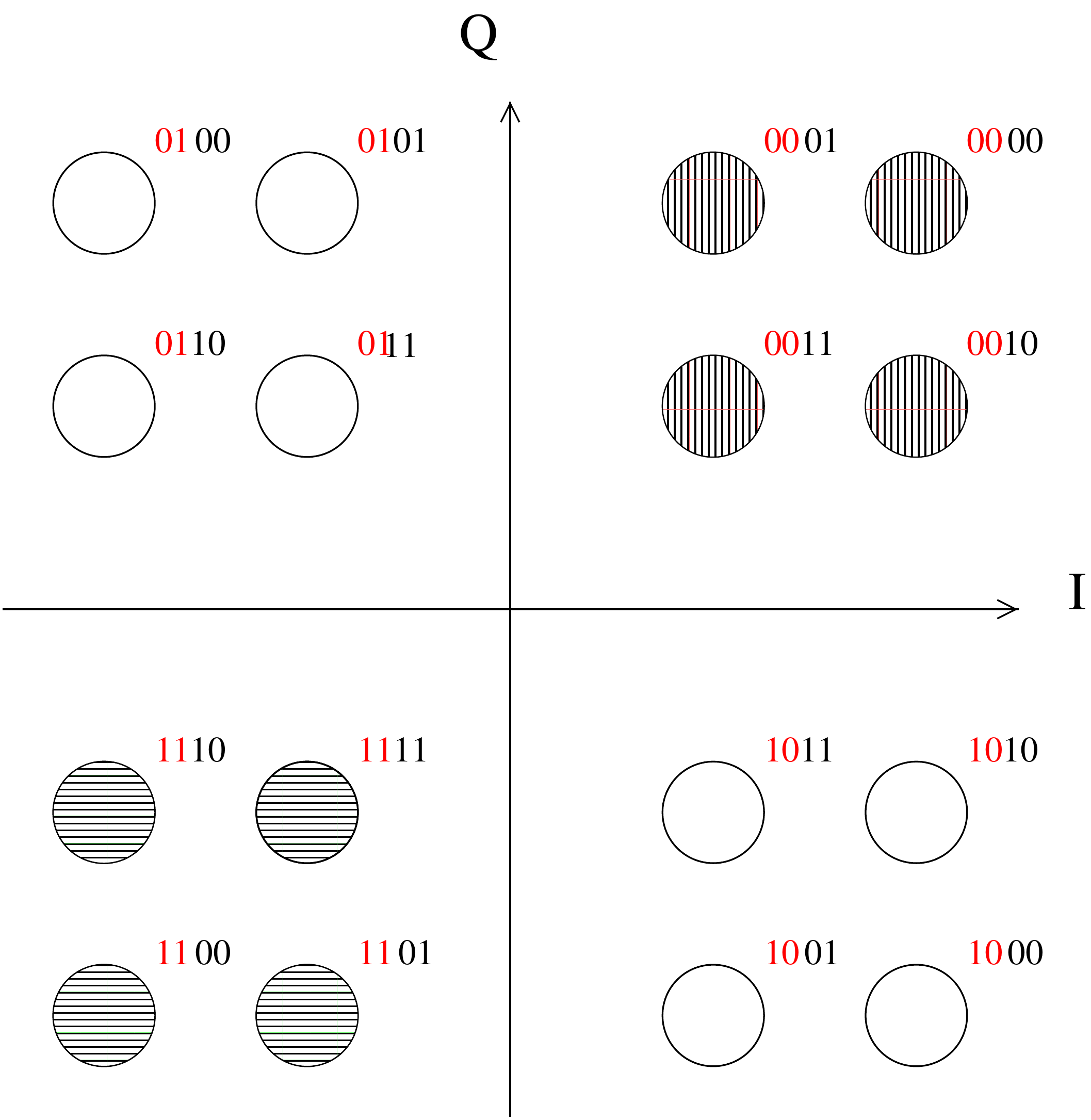}
\caption{Examples of $\chi_0$ (vertical lines) and $\chi_3$ (horizontal lines)}
\label{subset}
\end{figure}

The set $\chi_i$ depends on i and the positions of the bits involved in the stream. Figure~\ref{subset} shows an example of subsets for a 16-QAM with a particular mapping, where the stream uses bits in position 1 and 2 (in that case $k=2$, $i_1=1$ and $i_2=2$).

Then the conditional probability density function of $\mathbf{y}$ in (\ref{capacity_hm}) can be written
\setlength{\arraycolsep}{0.0em}
\begin{eqnarray}
p\left(\mathbf{y}|b_{i_1}=l_1(i),...,b_{i_k}=l_k(i) \right) &{}={}& 
\sum_{\mathbf{x} \in \chi_i} p(\mathbf{y}|\mathbf{x}) p(\mathbf{x}|\mathbf{x} \in \chi_i) \nonumber\\
&{=}& \frac{1}{|\chi_i|} \sum_{\mathbf{x} \in \chi_i} p(\mathbf{y}|\mathbf{x}) ,
\label{conditional_prob}
\end{eqnarray}
\setlength{\arraycolsep}{5pt}
where $|\chi_i|=2^{m-k}$ for all i. Moreover, the transition distribution $p(\mathbf{y}|\mathbf{x})$ for a Gaussian channel is 
\begin{eqnarray}
p(\mathbf{y}|\mathbf{x}) = \frac{1}{\pi N_0} \exp \left( -\frac{\| \mathbf{y}-\mathbf{x} \|^2}{N_0} \right).
\label{transition_distribution}
\end{eqnarray}

Using (\ref{conditional_prob}) and (\ref{transition_distribution}) in (\ref{capacity_hm}), we finally obtain the capacity for one stream (\ref{final_capacity_hm}). The capacity is an increasing function of $E_s/N_0$ and its value is less or equal to k the number of bits used in the stream. The positivity of the capacity results of its mathematical definition \cite{cove_thom_91}.
\setlength{\arraycolsep}{0.0em}
\begin{eqnarray}
C = k &{}-{}& \frac{1}{2^k\pi}\sum_{i=0}^{2^k-1} \int\limits_{-\infty}^{+\infty} \int\limits_{-\infty}^{+\infty} \left( \frac{1}{|\chi_i|}\sum_{\mathbf{x} \in \chi_i} \exp \left( \begin{Vmatrix} \mathbf{u}-\frac{\mathbf{x}}{\sqrt{N_0}} \end{Vmatrix}^2 \right) \right) \nonumber\\
&&{}\:\log_{2} \left( 1  + \frac{\displaystyle \sum_{\mathbf{x} \in \chi \setminus \chi_i} \exp \left( \begin{Vmatrix} \mathbf{u}-\frac{\mathbf{x}}{\sqrt{N_0}} \end{Vmatrix}^2 \right)}{\displaystyle \sum_{\mathbf{x} \in \chi_i} \exp \left( \begin{Vmatrix} \mathbf{u}-\frac{\mathbf{x}}{\sqrt{N_0}} \end{Vmatrix}^2 \right)}\right) \, \mathrm d\mathbf{u} 
\label{final_capacity_hm}
\end{eqnarray}
\setlength{\arraycolsep}{5pt}

\subsection{Case of non-uniform hierarchical modulation capacities}

We begin with few definitions before applying (\ref{final_capacity_hm}) to the hierarchical 16-QAM considered in DVB-SH.

Hierarchical modulations merge several streams in a same symbol. They often use non-uniform constellation. Non-uniform constellations are opposed to uniform constellations where the symbols are uniformly distributed. The constellation parameter is defined to describe non-uniform constellations. Figure~\ref{hm_principle} illustrates a non-uniform 16-QAM. The constellation parameter $\alpha$ is defined by $\alpha=d_h/d_l$ where $2d_h$ is the minimum distance between two constellation points carrying different HP bits and $2d_l$ is the minimum distance between any constellation points. Thus $\alpha$ verifies $\alpha \ge 1$ where $\alpha=1$ corresponds to the uniform 16-QAM. The DVB-SH standard \cite{sh} recommends two values for $\alpha$: 2 and 4.

\begin{figure}[!ht]
\centering
\includegraphics[width = 0.83\columnwidth]{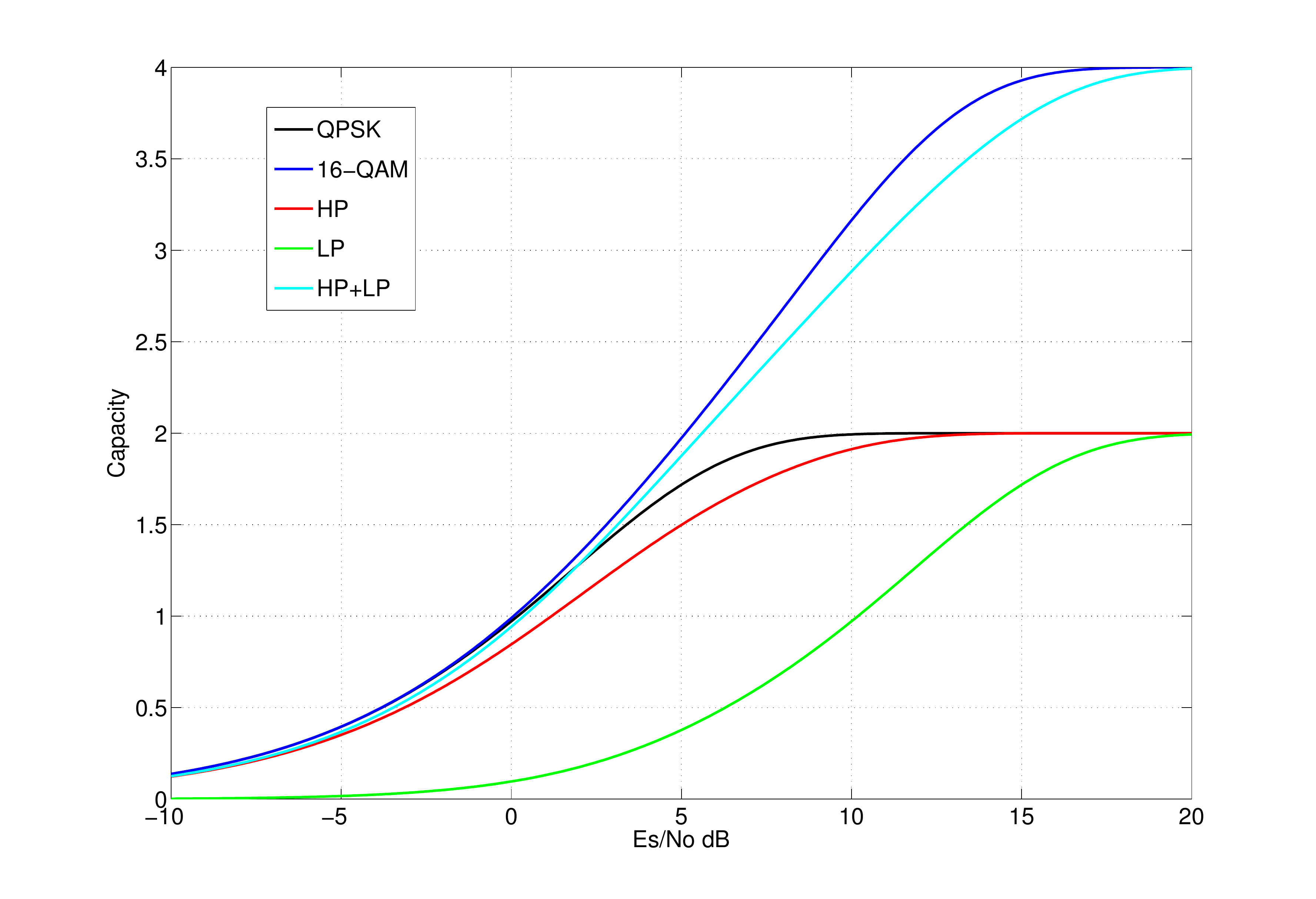}
\caption{16-QAM Hierarchical Modulation Capacity - $\alpha=2$}
\label{capacity_2}
\end{figure}

\begin{figure}[!ht]
\centering
\includegraphics[width = 0.83\columnwidth]{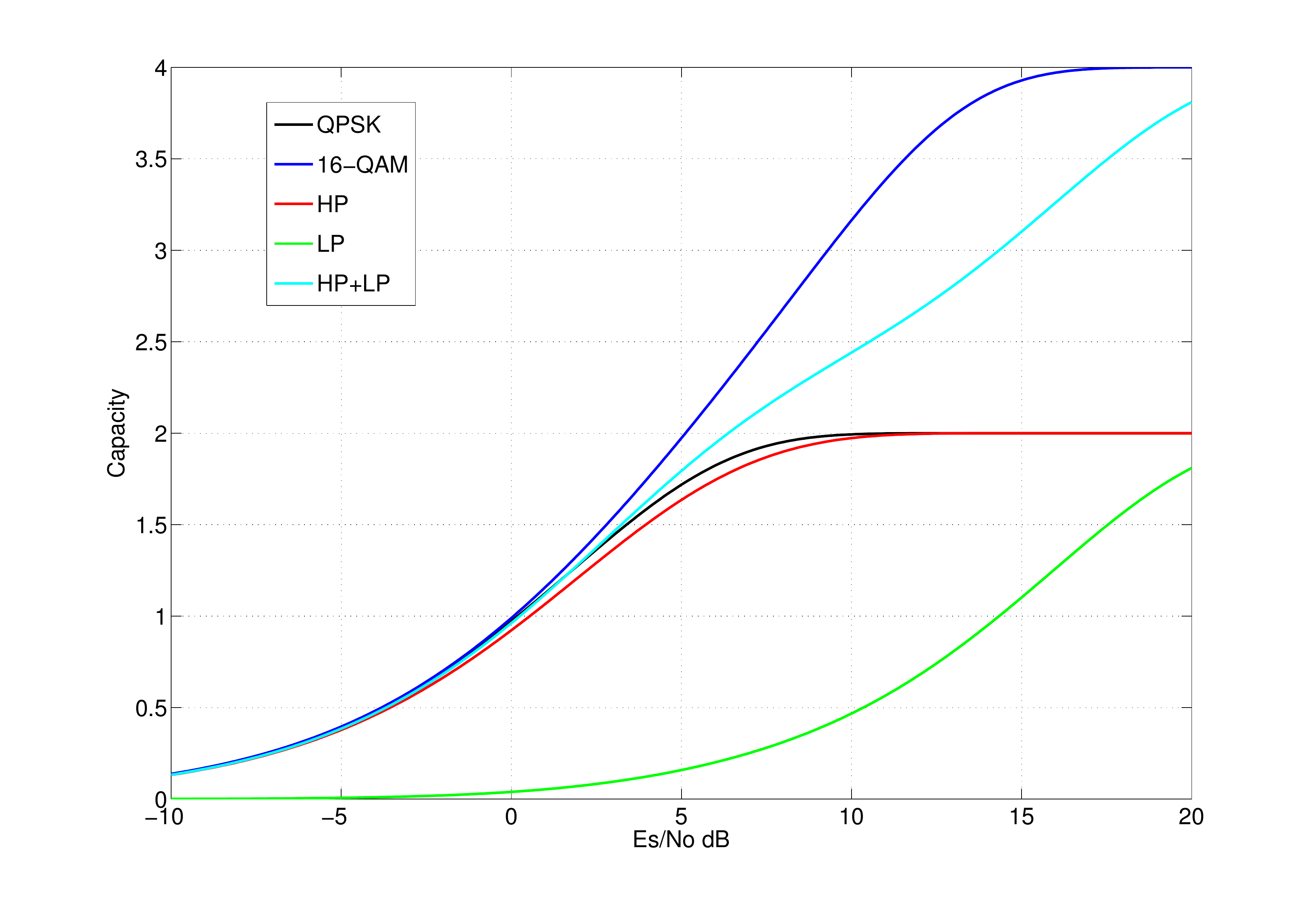}
\caption{16-QAM Hierarchical Modulation Capacity - $\alpha=4$}
\label{capacity_4}
\end{figure}

Back to the hierarchical modulation capacity, we suppose the HP stream uses the bits in position 1 and 2. The LP stream involves the remaining bits, i.e., bits 3 and 4. Figures~\ref{capacity_2} and \ref{capacity_4} present the 16-QAM capacity for the two values of $\alpha$ defined in the DVB-SH guidelines \cite{sh}.


When $\alpha$ grows, the constellation points in each quadrant become farer to the I and Q axes. Thus it is easier to decode the good quadrant and the capacity for the HP stream increases. However in the same quadrant, the points gets closer and the capacity for the LP stream decreases as it is harder to decode which symbol was sent. Concerning the HP stream capacity, it increases with $\alpha$ but nevertheless it is limited by the QPSK capacity as the same energy is used for 16 points instead of 4.

To conclude, we can observe that the global hierarchical modulation capacity (i.e. HP+LP) is always less than the 16-QAM capacity for any value of $\alpha > 1$. This result has been proved in \cite{caire} where the different streams use one bit. We can also note that the best total capacity is achieved when the constellation is close to the uniform constellation, but on the other hand performances for the HP stream are decreased.
 
%% file: result.tex
\section{Performance Evaluation for Hierarchical Modulations}\label{result}

We present in this part a method to get easily the spectrum efficiency and the required $E_s/N_0$ for a given targeted BER/PER for any modulation and coding rate without computing extensive simulations. In \cite{cnes} a fast coding/decoding performance evaluation method based on the mutual information computation is proposed and applied to time varying channel for QPSK modulation with very good prediction precision. The method developped hererafter based on channel capacity computation makes it possible to predict the performances of a coding scheme combined with any modulation and especially any hierarchical modulation.

\subsection{Principle}

\begin{figure}[!ht]
\centering
\includegraphics[width = 0.83\columnwidth]{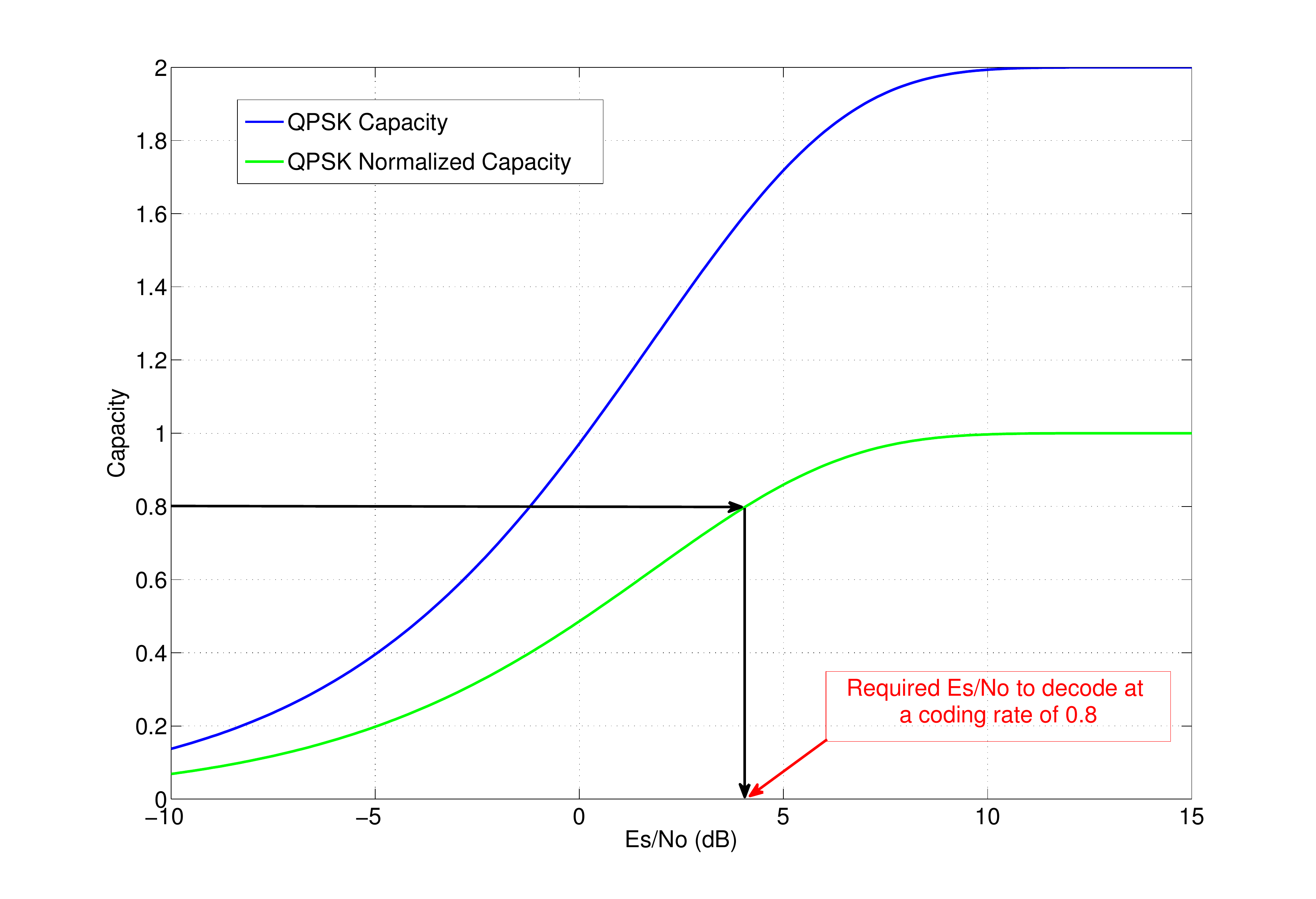}
\caption{Decoding threshold for a coding rate of 0.8}
\label{mean_capacity}
\end{figure}

Before applying the approach to a real code, we begin with an example using theoretical ideal codes achieving the channel capacity. The normalized capacity for a modulation is defined by $\overline{C}_{mod} = \frac{1}{m} C_{mod}$ where  $C_{mod}$ is the modulation's capacity and m denotes the number of bits per symbol. The $\overline{C}_{mod}$ function has some properties:
\begin{itemize}
\item It is an increasing function of $E_s/N_0$.
\item It verifies $0 \le \overline{C}_{mod} \le 1$.
\item It is equivalent to the notion of coding rate when using ideal codes.
\end{itemize}
Given a modulation and a coding rate, if we want to know at which $E_s/N_0$ the code is able to decode, we just need to inverse the normalized capacity function for that coding rate. Figure~\ref{mean_capacity} illustrates this example using a QPSK.

Let us now consider an actual waveform based on a real code. We can consider that the real code at a coding rate R works as an ideal code with rate $\tilde R$ where $\tilde R \ge R$. Determining $\tilde R$ requires a performance curve (BER or PER vs $E_s/N_0$) for \emph{one actual reference modulation} and \emph{coding scheme with code rate $R$}. The performance curve allows to estimate the behaviour of the code. We illustrate our method with an example. We would like to study the hierarchical modulation using an non-uniform 16-QAM ($\alpha=2$) in the DVB-SH standard at a target BER of $10^{-5}$. We use the 2/9 and 1/5-turbo codes for the HP and LP streams respectively. To determine $\tilde R$ and the decoding thresholds, the method works as follow:

\begin{enumerate}
\item Use the performance curve of the reference modulation with rate R to get the operating point $\left(E_s/N_0\right)_{ref}$ corresponding to the desired performance. In the DVB-SH guidelines \cite[Table 7.5]{sh}, we read: 
\begin{eqnarray*}
\text{For the coding rate 2/9, }\text{BER}_{\text{QPSK}} \left( -3.4 \text{dB} \right) = 10^{-5}\\
\text{For the coding rate 1/5, }\text{BER}_{\text{QPSK}} \left( -3.9 \text{dB} \right) = 10^{-5}
\end{eqnarray*}

\item Compute the normalized capacity for the reference modulation which corresponds to $\tilde R$: 
\begin{eqnarray*}
\text{For the HP stream, }\tilde R &=& \overline{C}_{\text{QPSK}} \left( -3.4 \text{dB} \right) \thickapprox 0.27\\
\text{For the LP stream, }\tilde R &=& \overline{C}_{\text{QPSK}} \left( -3.9 \text{dB} \right) \thickapprox 0.25
\end{eqnarray*} 

\item For the studied modulation, compute $E_s/N_0$ such as the normalized capacity at this SNR equals $\tilde R$:
\begin{eqnarray*}
\left( E_s/N_0 \right)_{HP} &=& \overline{C}_{\text{HP,}\alpha=2}^{-1} \left( \tilde R = 0.27 \right) = -2.7\text{dB}\\
\left( E_s/N_0 \right)_{LP} &=& \overline{C}_{\text{LP,}\alpha=2}^{-1} \left( \tilde R = 0.25 \right) = 6.2\text{dB}
\end{eqnarray*}
The DVB-SH guidelines provide the decoding thresholds for the HP and LP streams \cite[Figure 7.40]{sh}. Less the 0.3dB due to the pilots, we read -2.6dB and 6.2dB for the HP and LP streams respectively.

\item Finally the points $\left( \left(E_s/N_0\right)_{HP}, R_{HP} \times m \right)$ and $\left( \left(E_s/N_0\right)_{LP}, R_{LP} \times m \right)$ are plotted on the spectrum efficiency curve (e.g., Figure~\ref{eff_spec_2} and \ref{eff_spec_4}).
\end{enumerate}

These steps are repeated for all the coding rates. Our method makes two assumptions. First of all, it approximates the information rate by $R \times m$. This approximation is justified by the fact that the targeted performance ($\text{BER}=10^{-5}$) is very small and thus only hardly impact the useful information rate.

\begin{figure}[!ht]
\centering
\includegraphics[width = 0.83\columnwidth]{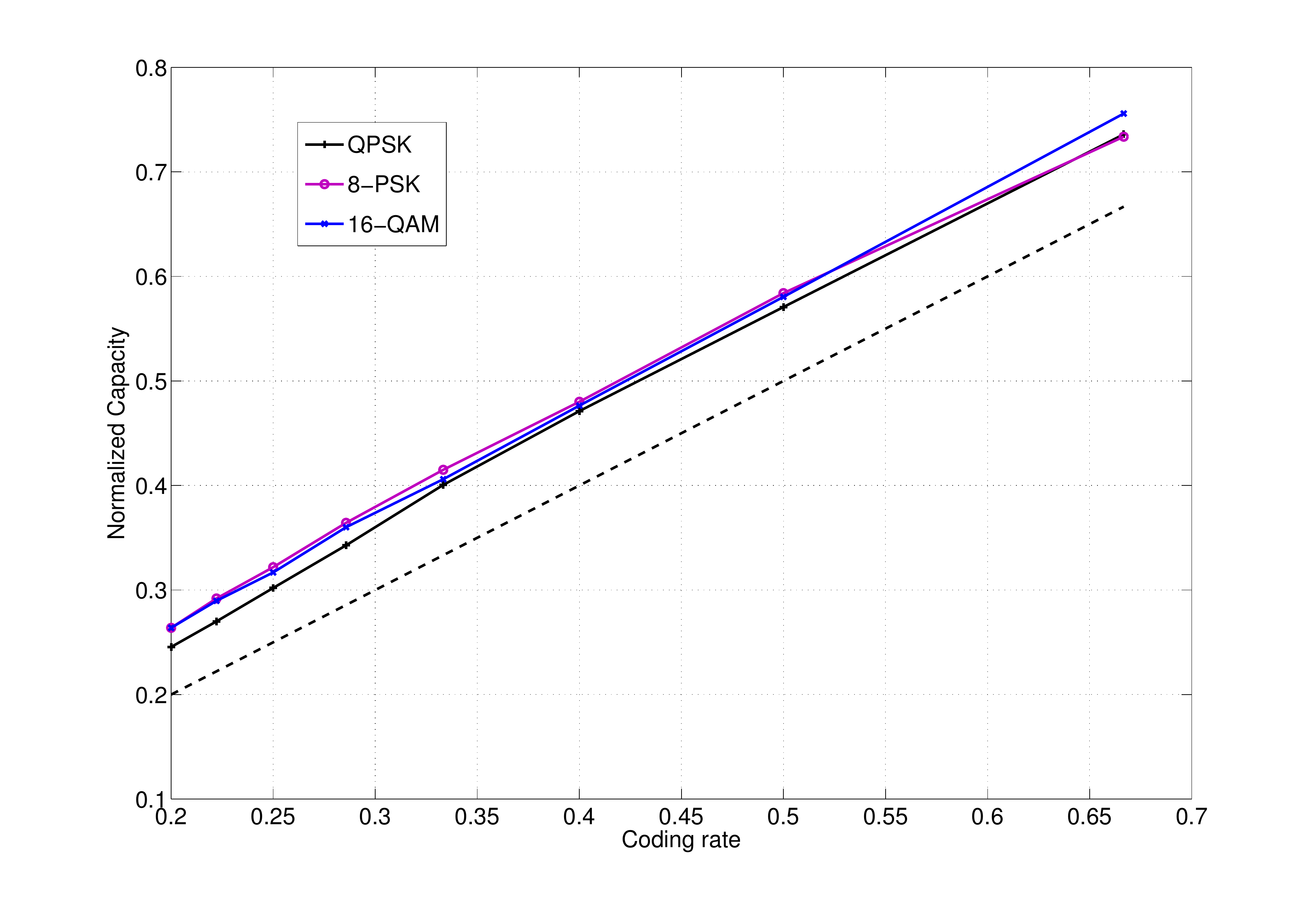}
\caption{DVB-SH: normalized capacities}
\label{mean_capacity_vs_coding}
\end{figure}

The second assumption is to suppose as in \cite{cnes} that \emph{the performance of the decoding only depends on the normalized capacity and not on the modulation as for ideal codes}. To validate this second assumption, Figure~\ref{mean_capacity_vs_coding} presents the normalized capacity in function of the coding rate for various modulations using the data of DVB-SH \cite[Table 7.5]{sh}.

\subsection{Application to DVB-SH}
In our study, we use the guidelines of DVB-SH \cite[Table 7.5]{sh} to get the reference curves and the operating points at a target BER of $10^{-5}$. These data correspond to a static receiver. The guidelines also provide all the reference numerical results for the hierarchical modulation \cite[Figure 7.40]{sh} where 0.3dB need to be removed due to the pilots. Thus it allows to evaluate the efficiency of our method. 

\begin{figure}[!ht]
\centering
\includegraphics[width = 0.82\columnwidth]{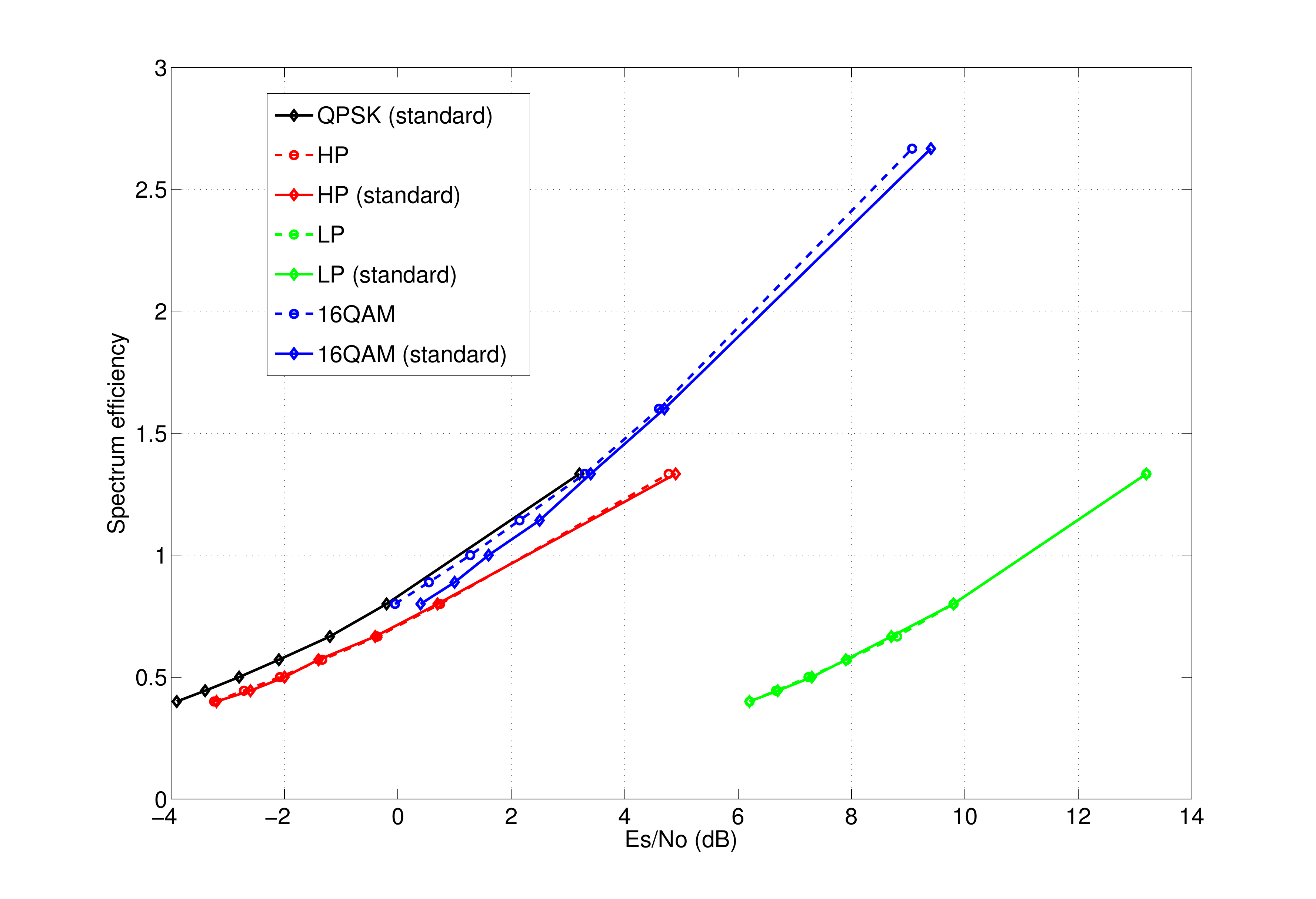}
\caption{DVB-SH spectrum efficiency, $\alpha=2$, $\text{BER}=10^{-5}$}
\label{eff_spec_2}
\end{figure}

\begin{figure}[!ht]
\centering
\includegraphics[width = 0.82\columnwidth]{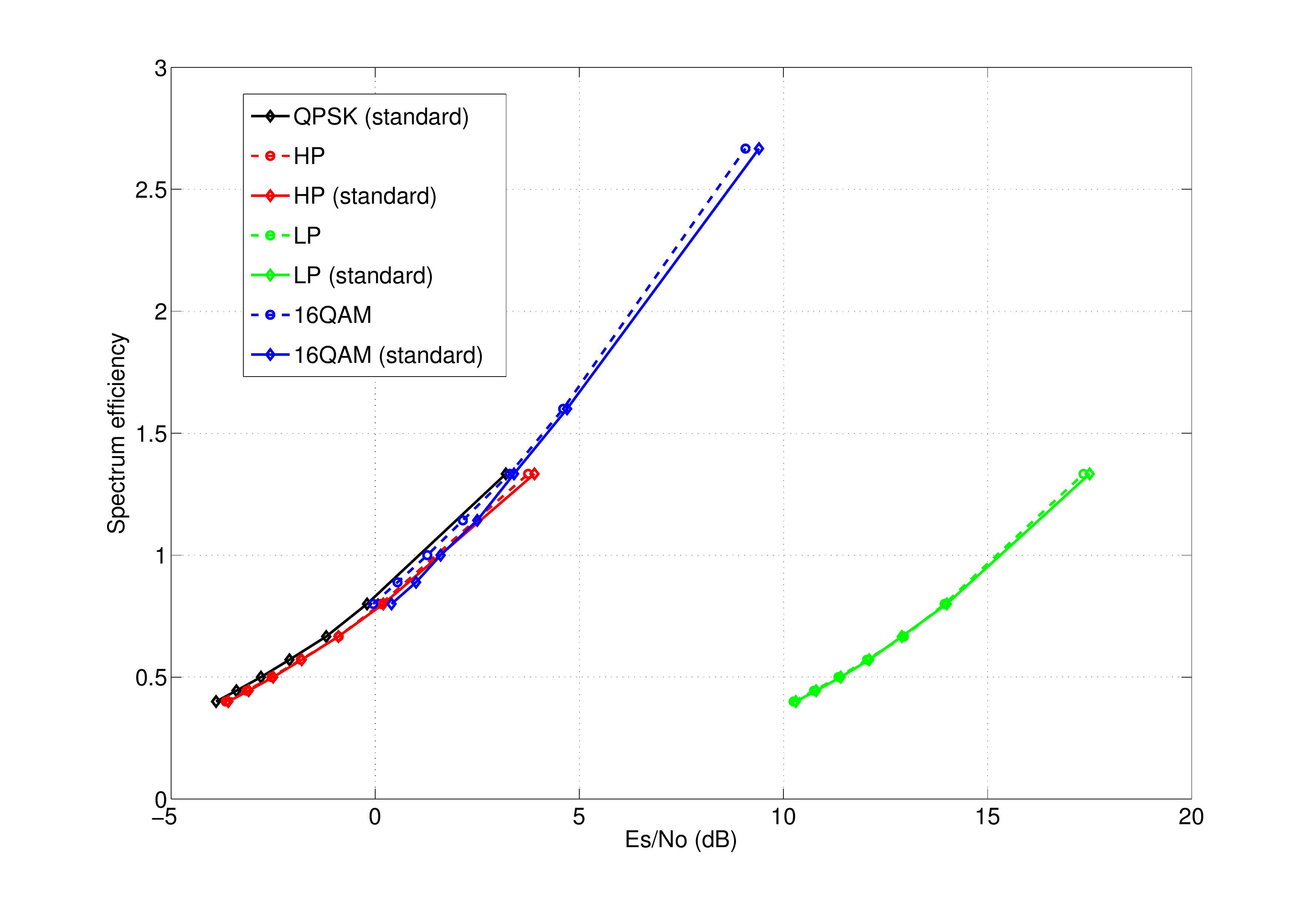}
\caption{DVB-SH spectrum efficiency, $\alpha=4$, $\text{BER}=10^{-5}$}
\label{eff_spec_4}
\end{figure}

The method described earlier is now applied to plot the spectrum efficiency curves as a function of required $E_s/N_0$ (target $\text{BER}=10^{-5}$). Figures~\ref{eff_spec_2} and \ref{eff_spec_4} present the results for $\alpha=2$ and $\alpha=4$ respectively. The reference results from the guidelines correspond to the standard curves. The results show a good precision and it does not require computing extensive simulations. Moreover, our work consolidates the fact that the capacity is a good metric to evaluate performance.

\subsection{DVB-SH: Additional Results}
We apply hereafter the method to compute the required $E_s/N_0$ in function of $\alpha$ or the coding rate.

\begin{figure}[!ht]
\centering
\includegraphics[width = 0.82\columnwidth]{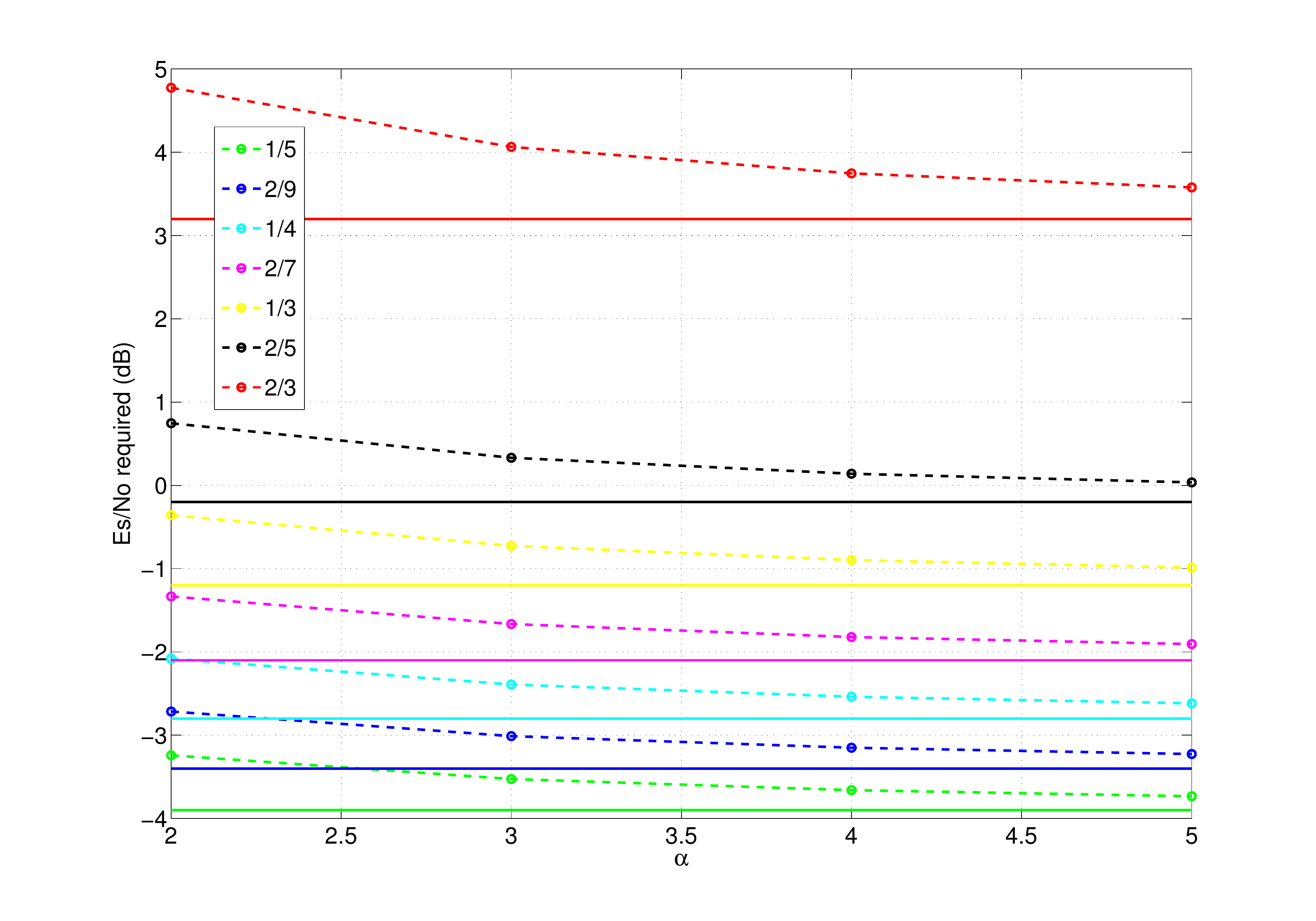}
\caption{Required $E_s/N_0$ function of $\alpha$, HP Stream, $\text{BER}=10^{-5}$}
\label{sgn_vs_alpha_hp}
\end{figure}

\begin{figure}[!ht]
\centering
\includegraphics[width = 0.82\columnwidth]{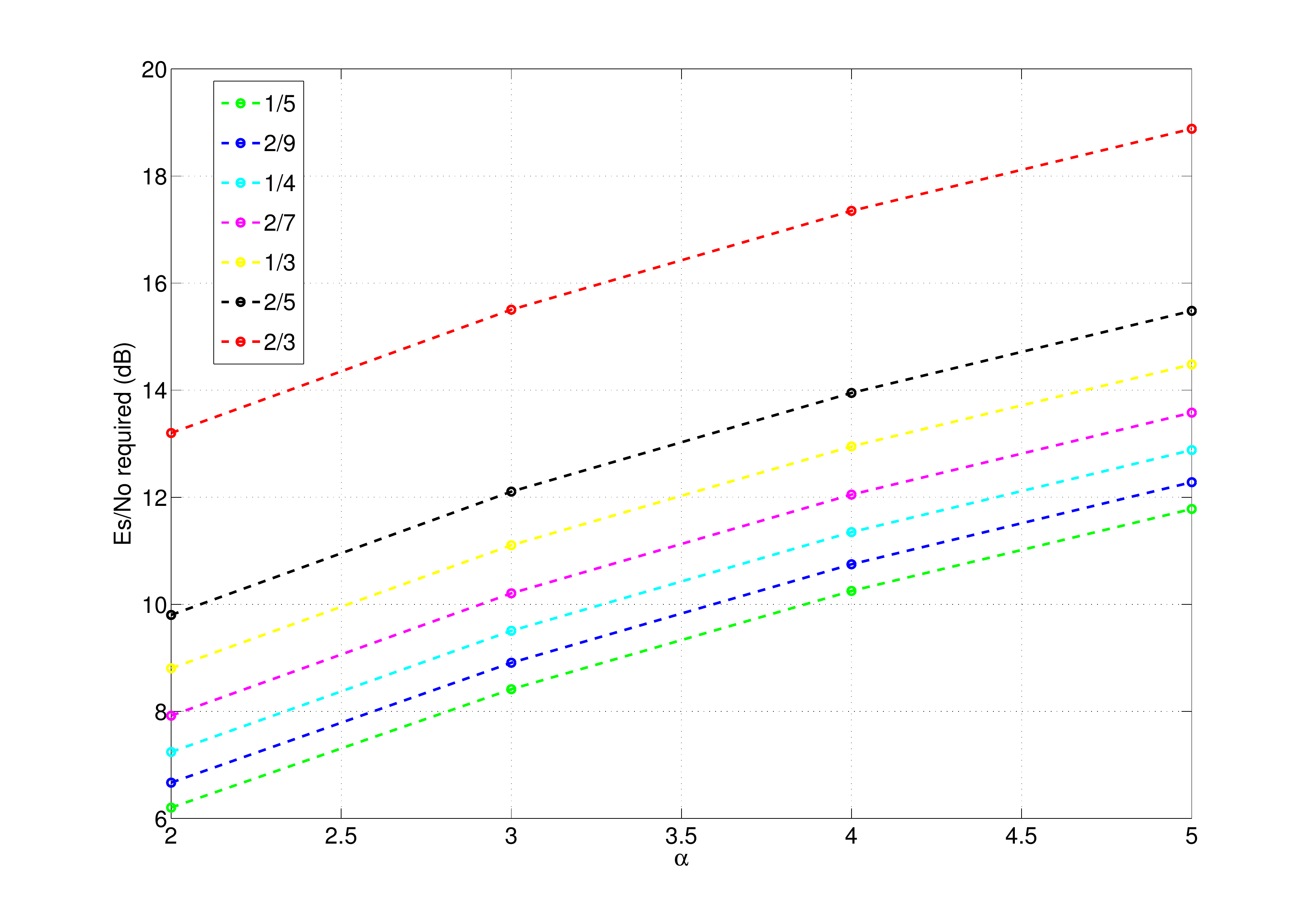}
\caption{Required $E_s/N_0$ function of $\alpha$, LP Stream, $\text{BER}=10^{-5}$}
\label{sgn_vs_alpha_lp}
\end{figure}

Figures~\ref{sgn_vs_alpha_hp} and \ref{sgn_vs_alpha_lp} present how the required $E_s/N_0$ varies with $\alpha$. For the LP stream, the required $E_s/N_0$ is an increasing function of $\alpha$, unlike the one for the HP stream who decreases. In fact the required $E_s/N_0$ for the HP stream tends to the required SNR of the QPSK modulation. These results are obvious when we look the modification of the constellation with $\alpha$. When $\alpha$ increases, the points in one quadrant become closer and the constellation is similar to a QPSK. It explains why the LP stream is harder to decode and requires a better SNR.

\begin{figure}[!ht]
\centering
\includegraphics[width = 0.82\columnwidth]{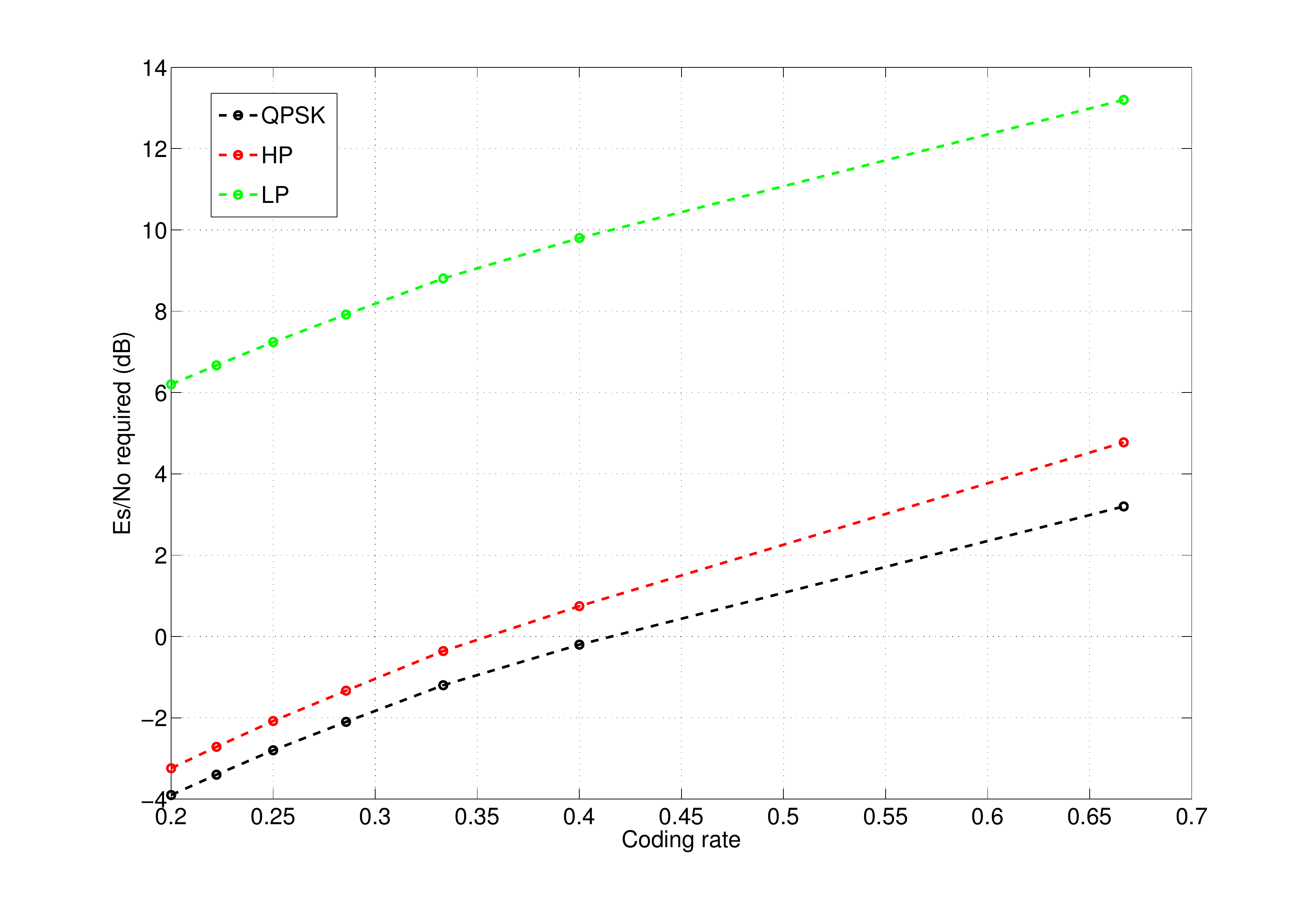}
\caption{Required $E_s/N_0$ function of the coding rate, $\alpha=2$}
\label{sgn_vs_coding_2}
\end{figure}

The last point concerns the variation of the required $E_s/N_0$ with the coding rate. It is presented on Figure~\ref{sgn_vs_coding_2} for $\alpha=2$. Obviously, if there is less redundancy, the SNR has to be higher to decode.

All these results are helpful during the dimensioning of a broadcast system. If the operator knows the SNR distribution of the receivers, it is possible to optimize the choice of the constellation parameter or the coding rate for each stream.

%% file: conclusion.tex
\section{Conclusion}\label{conclusion}

In this paper, we introduce a general method allowing to analyse the performance of hierarchical modulations. This method relies on the channel capacity which has been computed for any kind of constellation. It has been applied to DVB-SH in order to obtain the spectrum efficiency. The comparisons with reference simulation results show the good reliability of the method.

Future work will improve the comparison between hierarchical modulation and time sharing. We also plan to take into account the mobility in the context of DVB-SH. Finally we will apply our method to DVB-S2 and other standards including hierarchical modulation to complete the study.